\documentclass[aps, prb,showpacs,twocolumn, floatfix]{revtex4}
\usepackage{amsmath}
\usepackage{amssymb}
\usepackage{mathtools}
\usepackage{epsfig}
\usepackage{graphics}
\usepackage{color}
\usepackage{bm}
\usepackage{amssymb}
\usepackage{graphicx}
\begin{document}

\title{Does the Equivalence between Gravitational Mass and
Energy Survive for a Composite Quantum Body?}

\author{{\bf A.G. Lebed}$^*$}

\affiliation{Department of Physics, University of Arizona, 1118 E.
4th Street, Tucson, AZ 85721, USA}

\author{Correspondence should be addressed to A.G. Lebed:
lebed@physics.arizona.edu}

\begin{abstract}
We define passive and active gravitational mass operators of the
simplest composite quantum body - a hydrogen atom. Although they
do not commute with its energy operator, the equivalence between
the expectation values of passive and active gravitational masses
and energy is shown to survive for stationary quantum states. In
our calculations of passive gravitational mass operator, we take
into account not only kinetic and Coulomb potential energies but
also the so-called relativistic corrections to electron motion in
a hydrogen atom. Inequivalence between passive and active
gravitational masses and energy at a macroscopic level is
demonstrated to reveal itself as time dependent oscillations of
the expectation values of the gravitational masses for
superpositions of stationary quantum states. Breakdown of the
equivalence between passive gravitational mass and energy at a
microscopic level reveals itself as unusual electromagnetic
radiation, emitted by macroscopic ensemble of hydrogen atoms,
moved by small spacecraft with constant velocity in the Earth's
gravitational field. We suggest the corresponding experiment on
the Earth's orbit to detect this radiation, which would be the
first direct experiment where quantum effects in general
relativity are observed.
\end{abstract}

\maketitle

\section{Introduction}

Formulation of a successful quantum gravitational theory is
considered to be one of the most important problems in physics and
the major step towards the so-called "Theory of Everything". On
the other hand, fundamentals of general relativity and quantum
mechanics are so different that it is possible that these two
theories will not be united in the feasible future. In this
difficult situation, it seems to be important to suggest a
combination of quantum mechanics and some non-trivial
approximation of general relativity. In particular, this is
important in the case where such theory leads to meaningful
physical results, which can be experimentally tested.

A notion of gravitational mass of a composite body is known to be
non-trivial in general relativity and related to the following
paradoxes. If we consider a free photon with energy $E$ and apply
to it the so-called Tolman's formula for active gravitational mass
[1], we will obtain $m^g_a=2E/c^2$ (i.e., two times bigger value
than the expected one) [2]. If a photon is confined in a box with
mirrors, then we have a composite body at rest. In this case, as
shown in Ref. [2], we have to take into account a negative
contribution to $m^g_a$ from stress in the box walls to restore
the Einstein's equation, $m^g_a=E/c^2$. It is important that the
latter equation is restored only after averaging over time. A role
of the classical virial theorem in establishing the equivalence
between averaged over time active and passive gravitational masses
and energy is discussed in detail in Refs. [3,4] for different
types of classical composite bodies. In particular, for
electrostatically bound two bodies, it is shown that gravitational
field is coupled to a combination $3K+2U$, where $K$ is kinetic
and $U$ is the Coulomb potential energies. Since the classical
virial theorem states that the following time average is equal to
zero, $\bigl< 2K+U \bigl>_t = 0$, then we conclude that averaged
over time active and passive gravitational masses are proportional
to the total amount of energy [3,4],
 \begin{equation}
\bigl< m^g_{a,p} \bigl>_t = m_1 + m_2 + \bigl<3K + 2U \bigl>_t/c^2
= E/c^2,
\end{equation}
where $m_1$ and $m_2$ are bare masses of the above considered
bodies.

\section{Goal}

The main goal of our paper is to study a quantum problem about
passive [5-7] and active gravitational masses of a composite body.
As the simplest example, we consider a hydrogen atom. We claim
four main results in the paper. Our first result is that the
equivalence between passive and active gravitational masses and
energy survives at a macroscopic level for stationary quantum
states. In the calculations of passive gravitational mass
operator, we take into account both non-relativistic kinetic and
Coulomb potential energies and the so-called relativistic
corrections [8] to an electron motion in a hydrogen atom, whereas
in calculations of active gravitational mass we take into account
only non-relativistic kinetic and Coulomb potential energies. More
specifically, we show that the expectation values of passive and
active gravitational masses of the atom are equivalent to its
energy for stationary quantum states due to some mathematical
theorems. In the case of active gravitational mass, the
corresponding theorem is known as the quantum virial theorem [9],
whereas, in the case of passive gravitational mass, the
corresponding theorem is more complicated than that in Ref.[9]. In
fact the latter is an extension of the relativistic quantum virial
theorem [10] for the case of a particle with spin $\frac{1}{2}$.
We would like to draw attention
to the fact that the above-mentioned results are non-trivial.
Indeed, below we define passive and active gravitational mass
operators of an electron, $\hat m^g_p$ and $\hat m^g_a$,
respectively, in the post-Newtonian approximation to general
relativity. It is important that these operators occur not to
commute with electron energy operator, taken in the absence of the
field. Therefore, from the first point of view, it seems that the
equivalence between passive and active gravitational masses and
energy is broken. Nevertheless, using rather sophisticated
mathematical tools, we show that the expectation values of passive
and active gravitational mass operators are $<\hat m^g_p> = <\hat
m^g_a> = m_e + E_n / c^2$ for stationary quantum states in a hydrogen
atom, where $m_e$ is the bare electron mass, $E_n$ is the total
electron energy of n-th atomic energy level.

Our second result is that the equivalence between electron energy
and its passive and active gravitational masses is shown to be
broken for superpositions of stationary quantum states. More
strictly speaking, we demonstrate that there exist such quantum
states where the expectation values of energy are constant,
whereas the expectation values of passive and active gravitational
masses are oscillatory functions of time. Our third result is a
breakdown of the equivalence between passive gravitational mass
and energy at a microscopic level. It is a consequence of the fact
that passive electron gravitational mass operator, $\hat m^g_p$,
does not commute with its energy operator, taken in the absence of
the field. Therefore, an atom with a definite energy in the
absence of gravitational field, $E$, is not characterized by a
definite passive gravitational mass in an external gravitational
field. Passive gravitational mass is shown to be quantized and can
significantly differ from the value $E/c^2$. Our fourth result is
that we suggest how the above mentioned inequivalence can be
experimentally observed. In particular, we propose experimental
detection of electromagnetic radiation, emitted by macroscopic
ensemble of hydrogen atoms (in a real experiment - molecules),
supported by and moving with constant velocity in the Earth's
gravitational field, using small spacecraft or satellite. If such
experiment is done, to the best of our knowledge, it will be the
first direct experimental test of quantum effects in general
relativity. We stress, that so far only quantum effects in the
Newtonian variant of gravity, where general relativity corrections
are negligible, have been directly studied in the famous COW [11]
and ILL [12] experiments.

The most of the above mentioned results, related to passive
gravitational mass, have been recently published by us in Refs.
[5-7]. All results, related to active gravitational mass, and
results, related to the so-called relativistic corrections to
passive gravitational mass, are new and, to the best of our
knowledge, have not been published.

\section{Equivalence of the expectation values of passive
gravitational mass and energy for stationary states}

Let us use the standard weak field approximation to describe
spacetime outside the Earth [13,14],
\begin{eqnarray}
ds^2 = -\biggl(1 + \frac{2\phi}{c^2} \biggl)(cdt)^2 + \biggl(1 -
\frac{2 \phi}{c^2} \biggl) (dx^2 +dy^2+dz^2 ),
\nonumber\\
\phi = - \frac{GM}{R},
\end{eqnarray}
where $G$ is the gravitational constant, $c$ is the velocity of
light, $M$ is the Earth's mass, and $R$ is a distance from center of
the Earth. Then, in the local proper spacetime coordinates,
\begin{eqnarray}
&&x'=\biggl(1-\frac{\phi}{c^2} \biggl) x, \ y'=
\biggl(1-\frac{\phi}{c^2} \biggl) y,
\nonumber\\
&&z'=\biggl(1-\frac{\phi}{c^2} \biggl) z , \ t'=
\biggl(1+\frac{\phi}{c^2} \biggl) t,
\end{eqnarray}
the Schr\"{o}dinger equation for an electron motion in a hydrogen
atom can be approximately written in the following standard form:
\begin{equation}
i \hbar \frac{\partial \Psi({\bf r'},t')}{\partial t'} = \hat H
(\hat {\bf p'},{\bf r'}) \Psi({\bf r'},t')  ,
\end{equation}
where $\hat H (\hat {\bf p'},{\bf r'})$ is the standard
Hamiltonian. We stress that, in Eq.(4) and everywhere below, we
disregard all tidal effects (i.e., we do not differentiate
gravitational potential with respect to electron coordinates,
${\bf r}$ and ${\bf r'}$, corresponding to electron positions in
the center of mass coordinate systems). It is possible to show
that this means that we consider a hydrogen atom as a point-like
body and disregard all tidal terms in electron Hamiltonian, which are
usually very small and of the order of $(r_B/R_0)|\phi/c^2| \sim 10^{-17}|\phi/c^2| \sim
10^{-26}$ in the Earth's gravitational field. [Here $r_B$ is the
so-called Bohr's radius and $R_0$ is the Earth's radius.]

\subsection{Non-relativistic case}

Let us first consider non-relativistic Schr\"{o}dinger equation
for electron motion in a hydrogen atom, where we take into account only
kinetic and Coulomb potential energies:
\begin{eqnarray}
i \hbar \frac{\partial \Psi({\bf r'},t')}{\partial t'} = \hat H_0
(\hat {\bf p'},{\bf r'}) \Psi({\bf r'},t')  ,
\nonumber\\
 \ \hat H_0 (\hat {\bf p'},{\bf r'}) = m_e c^2 + \frac{\hat {\bf p'}^2}{2m_e}
 -\frac{e^2}{r'} ,
\end{eqnarray}
where $e$ is the electron charge, $r'$ is a distance between electron and proton,
and $\hat {\bf p'} = - i \hbar \partial /\partial {\bf r'}$ is
electron momentum operator in the local proper spacetime
coordinates. Below, we treat the weak gravitational field (2) as a
perturbation in inertial coordinate system, corresponding to
spacetime coordinates $(x,y,z,t)$ in Eq.(3) [3,4], and calculate
the corresponding Hamiltonian:
\begin{equation}
\hat H_0(\hat {\bf p},{\bf r}, \phi) = m_e c^2 + \frac{\hat {\bf
p}^2}{2m_e}-\frac{e^2}{r} + m_e  \phi + \biggl( 3 \frac{\hat {\bf
p}^2}{2 m_e} -2\frac{e^2}{r} \biggl) \frac{\phi}{c^2}.
\end{equation}
From Eq.(6), it is clear that the Hamiltonian can be rewritten in
the following form:
\begin{equation}
\hat H_0(\hat {\bf p},{\bf r}, \phi) = m_e c^2 + \frac{\hat {\bf
p}^2}{2m_e} -\frac{e^2}{r} + \hat m^g_e \phi \ ,
\end{equation}
where we introduce passive gravitational mass operator of an
electron:
\begin{equation}
\hat m^g_p  = m_e  + \biggl(\frac{\hat {\bf p}^2}{2m_e}
-\frac{e^2}{r}\biggl)/ c^2 + \biggl(2 \frac{\hat {\bf
p}^2}{2m_e}-\frac{e^2}{r} \biggl)/ c^2 \ ,
\end{equation}
which is proportional to its weight operator in the weak
gravitational field (2) [15]. Note that, in Eq.(8), the first term
corresponds to the bare electron mass, $m_e$, the second term
corresponds to the expected electron energy contribution to the
gravitational mass operator, whereas the third non-trivial term is
the virial contribution to passive gravitational mass operator. It
is possible to make sure [5,16] that Eqs.(7),(8) can be obtained
directly from the Dirac equation in a curved spacetime,
corresponding to the weak centrosymmetric gravitational field (2)
(see, for example, Eq.(3.24) in Ref. [17]), if we disregard all
tidal terms.

Here, we discuss some important consequence of Eqs.(7),(8). It is
crucial that the operator (8) does not commute with electron
energy operator, taken in the absence of gravitational field.
Therefore, it is not clear from the beginning that the equivalence
between electron passive gravitational mass and its energy exists.
To establish the equivalence at a macroscopic level, we consider a
macroscopic ensemble of hydrogen atoms with each of them being in
a stationary quantum state with a definite energy $E_n$. Then,
from Eq.(8), it follows that the expectation value of electron
passive gravitational mass operator per atom is
\begin{equation}
<\hat m^g_p > = m_e + \frac{ E_n}{c^2}  + \biggl< 2 \frac{\hat
{\bf p}^2}{2m_e}-\frac{e^2}{r} \biggl> /c^2 = m_e +
\frac{E_n}{c^2} ,
\end{equation}
where the third term in Eq.(9) is zero in accordance with the
quantum virial theorem [9]. Therefore, we conclude that the
equivalence between passive gravitational mass and energy survives
at a macroscopic level for stationary quantum states, if we
consider only pairings of non-relativistic kinetic and Coulomb
potential energies with an external gravitational field. Note that
an important difference between our result (9) and the
corresponding result in classical case [3,4] is that the
expectation value of passive gravitational mass corresponds to
averaging procedure over a macroscopic ensemble of hydrogen atoms,
whereas in classical case we average over time.

\subsection{Relativistic corrections}

In this section, we study a more general case, where the so-called
relativistic corrections to an electron motion in a hydrogen atom
are taken into account. As well known [8], there exist three
relativistic correction terms, which have different physical
meanings. The total Hamiltonian in the absence of gravitational
field can be written as:
\begin{equation}
\hat H (\hat {\bf p},{\bf r}) = \hat H_0 ( \hat {\bf p},{\bf r}) +
\hat H_1 (\hat {\bf p},{\bf r}) ,
\end{equation}
where
\begin{equation}
\hat H_1 (\hat {\bf p},{\bf r}) = \alpha \hat {\bf p}^4 + \beta
\delta^3 ({\bf r}) + \gamma \frac{\hat {\bf S} \cdot \hat {\bf L}
}{r^3} ,
\end{equation}
with the parameters $\alpha$, $\beta$, and $\gamma$ being:
\begin{equation}
\alpha = - \frac{1}{8 m_e^3 c^2}, \ \beta=\frac{\pi e^2
\hbar^2}{2m_e^2c^2}, \ \gamma = \frac{e^2}{2 m_e^2 c^2}.
\end{equation}
[Here, $\delta^3 ({\bf r}) = \delta (x) \delta (y) \delta (z)$ is
a three dimensional Dirac's delta-function.] Note that the first
contribution in Eq.(11) is called the kinetic term, the second one
is the so-called Darwin's term, and the third one is the
spin-orbital interaction, where $\hat {\bf L} = - i \hbar [{\bf r}
\times
\partial / \partial {\bf r}]$ is electron angular momentum
operator. In the presence of the weak gravitational field (2), the
Schr\"{o}dinger equation for an electron motion in the local
proper spacetime coordinates (3) can be approximately written as:
\begin{equation}
i \hbar \frac{\partial \Psi({\bf r'},t')}{\partial t'} = [\hat
H_0(\hat {\bf p'},{\bf r'})+ \hat H_1(\hat {\bf p'},{\bf r'})]
\Psi ({\bf r',t'}).
\end{equation}
[Note that, as discussed above, we disregard everywhere all tidal
effects.]

By means of the coordinates transformation (3), the corresponding
Hamiltonian in inertial coordinate system (x,y,z,t) can be
expressed as:
\begin{eqnarray}
\hat H(\hat {\bf p},{\bf r}, \phi)= [\hat H_0(\hat {\bf p}, {\bf
r}) + \hat H_1 (\hat {\bf p},{\bf r})] \biggl(1 + \frac{\phi}{c^2}
\biggl)
\nonumber\\
+ \biggl(2 \frac{\hat {\bf p}^2}{2 m_e}-\frac{e^2}{r} + 4 \alpha
\hat {\bf p}^4 + 3 \beta \delta^3({\bf r}) + 3 \gamma \frac{\hat
{\bf S} \cdot \hat {\bf L} }{r^3} \biggl) \frac{\phi}{c^2} .
\end{eqnarray}
For the Hamiltonian (14), passive gravitational mass operator of
an electron can be written in more complicated than Eq.(8) form:
\begin{eqnarray}
&&\hat m^g_p = m_e + \biggl( \frac{\hat {\bf p}^2}{2m_e} -
\frac{e^2}{r} + \alpha \hat {\bf p}^4 + \beta \delta^3 ({\bf r}) +
\gamma \frac{\hat {\bf S} \cdot \hat {\bf L} }{r^3} \biggl)/c^2
\nonumber\\
&&+ \biggl(2 \frac{\hat {\bf p}^2}{2 m_e}-\frac{e^2}{r} + 4 \alpha
\hat {\bf p}^4 + 3 \beta \delta^3 ({\bf r}) + 3 \gamma \frac{\hat
{\bf S} \cdot \hat {\bf L} }{r^3} \biggl)/c^2 .
\end{eqnarray}
Let us consider again a macroscopic ensemble of hydrogen atoms
with each of them being in a stationary quantum state with a
definite energy $E'_n$, where $E'_n$ takes into account the
relativistic corrections (11) to electron energy. In this case,
the expectation value of the electron mass operator (15) per atom
can be written as:
\begin{eqnarray}
&&<\hat m^g_p > = m_e + \frac{ E'_n}{c^2}
\nonumber\\
&&+ \biggl<2 \frac{\hat {\bf p}^2}{2 m_e}-\frac{e^2}{r} + 4 \alpha
\hat {\bf p}^4 + 3 \beta \delta^3 ({\bf r}) + 3 \gamma \frac{\hat
{\bf S} \cdot \hat {\bf L} }{r^3} \biggl>/c^2 .
\end{eqnarray}

Below, we show that the expectation value of the third term in
Eq.(16) is zero and, therefore, the Einstein's equation, related
the expectation value of passive gravitational mass and energy, can be
applied to stationary quantum states. Here, we define the
so-called virial operator [9],
\begin{equation}
\hat G = \frac{1}{2} (\hat {\bf p}{\bf r} +{\bf r} \hat {\bf p}) ,
\end{equation}
and write the standard equation of motion for its expectation
value:
\begin{equation}
\frac{d}{dt} \biggl< \hat G \biggl> = \frac{i}{\hbar} \biggl<
[\hat H_0(\hat {\bf p},{\bf r}) + H_1(\hat {\bf p},{\bf r}), \hat
G] \biggl> ,
\end{equation}
where $[\hat A, \hat B]$, as usual, stands for a commutator of two
operators, $\hat A$ and $\hat B$. If we consider a stationary
quantum state with a definite energy, $E'_n$, then the derivative
$d<\hat G>/dt$ in Eq.(18) has to be zero and, thus,
\begin{equation}
\biggl< [\hat H_0(\hat {\bf p},{\bf r})+ H_1(\hat {\bf p},{\bf
r}), \hat G] \biggl> = 0 ,
\end{equation}
where the Hamiltonians $\hat H_0(\hat {\bf p},{\bf r})$ and $\hat
H_1(\hat {\bf p},{\bf r})$ are defined by Eqs.(5),(11). By means
of rather lengthy but straightforward calculations it is possible
to show that
\begin{eqnarray}
&&\frac{[\hat H_0(\hat {\bf p},{\bf r}), \hat G]}{-i \hbar}= 2
\frac{\hat {\bf p}^2}{2m_e}-\frac{e^2}{r} , \ \ \frac{[\alpha \hat
{\bf p}^4,\hat G]}{-i \hbar} = 4 \alpha \hat {\bf p}^4 ,
\nonumber\\
&&\frac{[\beta \delta^3 ({\bf r}), \hat G]}{-i \hbar} = 3 \beta
\delta^3 ({\bf r}), \ \ \frac{1}{-i \hbar} \biggl[ \gamma
\frac{\hat {\bf S} \cdot \hat {\bf L}}{r^3} , \hat G \biggl] = 3
\gamma \frac{\hat {\bf S} \cdot \hat {\bf L}}{r^3} ,
\end{eqnarray}
where we take into account the following equality: $x_i \frac{d
\delta(x_i)}{d x_i} = - \delta (x_i)$. As directly follows from
Eqs.(19),(20),
\begin{equation}
\biggl<2 \frac{\hat {\bf p}^2}{2 m_e}-\frac{e^2}{r} + 4 \alpha
\hat {\bf p}^4 + 3 \beta \delta^3 ({\bf r}) + 3 \gamma \frac{\hat
{\bf S} \cdot \hat {\bf L} }{r^3} \biggl> =0,
\end{equation}
and, therefore, Eq.(16) can be rewritten in the Einstein's form:
\begin{equation}
<\hat m^g_p > = m_e + \frac{E'_n}{c^2} .
\end{equation}
[Note that Eq.(21) extends the so-called relativistic quantum
virial theorem [10], derived for spinless particles, to the case of
particles with spin $\frac{1}{2}$.]

It is important that Eq.(22) directly establishes the equivalence
between the expectation value of electron passive gravitational
mass and its energy in a hydrogen atom, including the relativistic
corrections, for the E\"{o}tv\"{o}s' type of experiments [13]. We
speculate that such equivalence exists also for more complicated
quantum systems, including many-body systems with arbitrary
interactions of particles. These reveal and establish the physical
meaning of a coupling of a macroscopic quantum test body with a
weak gravitational field.

\section{Inequivalence between passive gravitational mass and energy
for superpositions of stationary states}

In the previous section, we have shown that the expectations
values of passive gravitational mass and energy are equivalent to
each other in stationary quantum states. Here, we investigate if
such equivalence survives for superpositions of stationary quantum
states. For this purpose, we consider the simplest superposition
of the ground and first excited $s$-wave states in a hydrogen
atom, where electron wave function has the the following form:
\begin{equation}
\Psi_{1,2}(r,t) = \frac{1}{\sqrt{2}} \bigl[ \Psi_1(r) \exp(-iE_1t)
+ \Psi_2(r) \exp(-iE_2t) \bigl].
\end{equation}
It is important that wave function (23) is characterized by the
time independent expectation value of energy, $<E> = (E_1+E_2)/2$.
Nevertheless, the expectation value of passive gravitational mass
operator (8) occurs to be the following time dependent oscillatory
function:
\begin{equation}
<\hat m^g_p> = m_e + \frac{E_1+E_2}{2 c^2} + \frac{V_{1,2}}{c^2}
\cos \biggl[ \frac{(E_1-E_2)t}{\hbar} \biggl],
\end{equation}
where $V_{1,2}$ is a matrix element of the virial operator:
\begin{equation}
V_{1,2}= \int \Psi^*_1(r) \biggl(2 \frac{\hat {\bf
p}^2}{2m_e}-\frac{e^2}{r} \biggl) \Psi_2(r) d^3{\bf r} .
\end{equation}
Note that the above obtained result holds both for
non-relativistic passive gravitational mass operator (8) and for
the operator (15), which takes into account the so-called
relativistic corrections (11). Therefore, we make a conclusion
that the oscillations of passive gravitation mass (24) directly
demonstrate inequivalence of passive gravitational mass and energy
at a macroscopic level at any given moment of time.

Let us discuss a relative magnitude of the oscillations (24). By
using actual numbers for a hydrogen atom, it is possible to obtain
the following numerical value of the matrix element of the virial
operator: $V_{1,2}=5.7 \ eV$. Since $m_e c^2 \simeq 0.5 \ MeV$ and
$m_p \simeq 1800 \ m_e$, we can come to the conclusion that the
oscillations (24) are weak but not negligible: $\delta m_e / m_e
\sim 10^{-5}$ and $\delta m_e / m_p \sim 10^{-8}$. They correspond
to the following angular and linear frequencies: $\omega_{1,2}
\simeq 1.6 \times 10^{16} Hz$ and $\nu \simeq 2.5 \times 10^{15}
Hz$, respectively. We hope that the above mentioned oscillations
of passive gravitational mass are experimentally measured, despite
the fact that the quantum state (23) decays with time.

On the other hand, if we average the oscillations (24) over time, we
obtain the modified equivalence principle between the averaged over time
expectation value of passive gravitational mass and the expectation
value of energy in the following form:
\begin{equation}
<< \hat m^g_p >>_t = m_e + (E_1+E_2)/2c^2 = E/c^2 .
\end{equation}
We pay attention that the physical meaning of averaging procedure in Eq.(26)
is completely different from that for classical time averaging procedure
(1) and does not have the corresponding classical analogs.

\section{Breakdown of the equivalence between passive
gravitational mass and energy at a microscopic level}

In this section, we study how non-commutation of passive
gravitational mass operators (8),(15) and the corresponding energy
operators, taken in the absence of gravitational field, results in
a breakdown of the equivalence between passive gravitational mass
and energy. This conclusion does not depend on the relativistic
corrections (11), therefore, for certainty, below we consider
passive gravitational mass operator in the form of Eq.(8). The
physical meaning of the above mentioned breakdown is that an
electron in its ground state with a definite energy, $m_ec^2+
E_1$, is not characterized by a definite passive gravitational
mass and, thus, measurements of the mass can give values, which
are not related to electron energy by the Einstein's equation,
$m^g_p \neq m_e +E_1/c^2$. As we show below, the passive electron
gravitational mass values in a hydrogen atom are quantized:
$m^g_p=m_e+E_n/c^2$, where $E_n$ is energy corresponding to n-th
energy level.

\subsection{First thought experiment}

Here, we describe the first thought experiment, illustrating
inequivalence of energy and passive gravitational mass at a
microscopic level. Suppose that we create quantum state of a
hydrogen atom with a definite energy in the absence of a
gravitational field and then adiabatically switch on the
gravitational field (2). More specifically, at $t \rightarrow -
\infty$ (i.e., in the absence of gravitational field), a hydrogen
atom is in its ground state with wave function,
\begin{equation}
\Psi_1(r,t) = \Psi_1(r) \exp(-im_ec^2t/\hbar -iE_1t/\hbar) \ ,
\end{equation}
whereas, in the vicinity of $t=0$ [i.e., in the presence of the
gravitational field (2), it is characterized by the following wave
function:
\begin{equation}
\Psi(r,t) = \sum^{\infty}_{n = 1} a_n (t) \Psi_n(r)
\exp(-im_ec^2t/\hbar -i E_n t/\hbar) \ .
\end{equation}
[Here, $\Psi_n(r)$ is a normalized electron wave function in a
hydrogen atom in the absence of gravitational field, corresponding
to energy $E_n$ [18].]

As follows from Eqs.(7),(8), adiabatically switched on gravitational field corresponds
to the following time-dependent small
perturbation:
\begin{eqnarray}
\hat U ({\bf r},R,t) = \phi(R)  \biggl[m_e + \biggl(\frac{\hat
{\bf p}^2}{2m_e}-\frac{e^2}{r}\biggl)/c^2
\nonumber\\
+\biggl(2 \frac{\hat {\bf p}^2}{2m_e}-\frac{e^2}{r} \biggl)/ c^2 \biggl] \exp(\lambda t) ,
\end{eqnarray}
where $\lambda \rightarrow 0$ [19]. The standard calculations by
means of the time-dependent quantum mechanical perturbation theory
[8] give the following results:

\begin{equation}
a_1(t)=\exp \Big[ - \frac{i \phi(R) m_e c^2 t + i \phi(R) E_1  t }{c^2
\hbar} \Big] \  ,
\end{equation}

\begin{equation}
a_n(0)=  - \frac{\phi(R)}{c^2} \frac{V_{n,1}}{E_n-E_1} \  , \ n \neq 1 \ ,
\end{equation}
where
\begin{equation}
V_{n,1}= \int \Psi^*_n(r) \biggl( 2 \frac{{\bf \hat p}^2}{2 m_e} -
\frac{e^2}{r} \biggl) \Psi_1(r) d^3 {\bf r} \ .
\end{equation}
[Note that the perturbation (29) is characterized by the following selection
rule. Electron from $1S$ ground state of a hydrogen atom can be excited
only into $nS$ excited state.]

Let us discuss Eqs.(30)-(32). It is important that Eq.(30)
corresponds to the well-known red shift of atomic ground state
energy $E_1$ in the gravitational field (2). On the other hand,
Eq.(31) demonstrates that there is a finite probability,
\begin{equation}
P_n = |a_n(0)|^2 = \Big[ \frac{\phi(R)}{c^2} \Big]^2 \
\Big(  \frac{V_{n,1}}{E_n-E_1} \Big)^2 \ , \ n \neq 1,
\end{equation}
that, at $t=0$, electron occupies n-th energy level. In fact, this
means that measurements of gravitational mass (8) in a quantum
state with definite energy (27) give the following quantized
values:
\begin{equation}
m^g_p (n) = m_e + E_n / c^2 \ ,
\end{equation}
with the probabilities (33) for $n \neq 1$. Note that, although
the probabilities (33) are quadratic with respect to gravitational
potential and, thus, small,  the corresponding changes of
gravitational mass (34) are large and of the order of $\alpha^2
m_e$, where $\alpha$ is the fine structure constant. It is
important that the excited levels of a hydrogen atom spontaneously
decay, therefore, one can detect the above discussed quantization
law of gravitational mass (34) by measuring electromagnetic
radiation, emitted by a macroscopic ensemble of hydrogen atoms.

\subsection{Second thought experiment}

Let us discuss the second thought experiment, which directly
demonstrates inequivalence between energy and passive
gravitational mass at a microscopic level. Suppose that, at $t=0$,
we create a ground state wave function of a hydrogen atom,
corresponding to the absence of gravitational field [see Eq.(27)].
Then, in the presence of the gravitational field (2), the wave
function (27) is not anymore a ground state of the Hamiltonian
(7),(8), where we treat gravitational field as a small
perturbation in inertial coordinate system [3,4,17]. It is
important that for an inertial observer, in accordance with
Eqs.(3), a general solution of the Schr\"{o}dinger equation,
corresponding to the Hamiltonian (7),(8), can be written as:
\begin{eqnarray}
\Psi(r,t) = (1-\phi/c^2)^{3/2} \sum^{\infty}_{n = 1} a_n \Psi_n
[(1- \phi/c^2)r] \nonumber\\
\exp[-i m_e c^2 (1+\phi/c^2) t/\hbar] \nonumber\\
\exp[-i E_n(1+\phi/c^2) t/\hbar] \ .
\end{eqnarray}
We pay attention that wave function (35) is a series of
eigenfunctions of passive gravitational mass operator (8), if we
take into account only linear terms with respect to small
parameter $\phi/c^2$. [Here, factor $1-\phi/c^2$ is due to a
curvature of space, whereas the term $E_n(1+\phi/c^2)$ represents
the famous red shift in gravitational field. We also pay attention
that the wave function (35) contains a normalization factor
$(1-\phi/c^2)^{3/2}$.]

In accordance with the basic principles of the quantum mechanics,
probability that, at $t>0$, an electron occupies excited state
with energy $m_e c^2(1+\phi/c^2) + E_n(1+\phi/c^2)$ is
\begin{eqnarray}
P_n = |a_n|^2, \ a_n = \int \Psi^*_1(r) \Psi_n [(1-\phi/c^2)r] d^3
{\bf r}
\nonumber\\
= - ( \phi/c^2) \int \Psi^*_1(r) r \Psi'_n(r)
d^3 {\bf r}.
\end{eqnarray}
[Note that it is possible to demonstrate that for $a_1$ in Eq.(36)
a linear term with respect to gravitational potential, $\phi$, is
zero, which is a consequence of the quantum virial theorem.]
Taking into account that the Hamiltonian is a Hermitian operator,
it is possible to show that for $n \neq 1$:
\begin{equation}
\int \Psi^*_1(r) r \Psi'_n(r) d^3 {\bf r} = V_{n,1}/ (E_n - E_1) ,
\end{equation}
where $V_{n,1}$ is a matrix element of the virial operator given
by Eq.(32).

Let us discuss Eqs.(36),(37). We stress that they directly
demonstrate that there is a finite probability,
\begin{equation}
P_n = |a_n|^2 = \Big[ \frac{\phi(R)}{c^2} \Big]^2 \ \Big(
\frac{V_{n,1}}{E_n-E_1} \Big)^2 \ , \ n \neq 1,
\end{equation}
that, at $t>0$, an electron occupies n-th ($n \neq 1$) energy
level, which breaks the expected Einstein's equation, $m^g_p=m_e +
E_1/c^2$. In fact, this means that quantum measurement of passive
gravitational mass [i.e., weight in the gravitational field (2)]
in a quantum state with a definite energy (27) gives the quantized
values [see Eq.(34)], corresponding to the probabilities
(33),(38), which are equal. [Note that, as it follows from quantum
mechanics, we have to calculate wave function (35) in a linear
approximation with respect to small parameter $\phi/c^2$ to obtain
probabilities (38), which are proportional to $(\phi/c^2)^2$. A
simple analysis shows that inclusion in Eq.(35) terms of the order
of $(\phi/c^2)^2$ would change electron passive gravitational mass
of the order of  $(\phi / c^2) m_e \sim 10^{-9} m_e$, which is
much smaller than the typical distance between the quantized
values in Eq. (34), $\delta m^g_p \sim \alpha^2 m_e \sim 10^{-4}
m_e$.] We pay attention that small values of probabilities
(33),(38), $P_n \sim 10^{-18}$, do not contradict to the existing
E\"{o}tvos type measurements [13], which have confirmed the
equivalence principle with the accuracy of the order of $\delta m/
m \sim 10^{-12}-10^{-13}$. As we mentioned in the previous
section, for our case, it is crucial that the excited levels of a
hydrogen atom spontaneously decay with time, therefore, one can
detect the quantization law (34) by measuring electromagnetic
radiation, emitted by a macroscopic ensemble of hydrogen atoms.
The above mentioned optical method is much more sensitive than the
E\"{o}tvos type measurements and we, therefore, believe that it
will allow to detect the breakdown of the equivalence between
passive gravitational mass and energy, revealed in the paper.

\section{Suggested realistic experiment}

\subsection{Hamiltonian}

Let us consider a realistic experiment, which can be done on the
Earth's orbit to detect photons, emitted by a macroscopic ensemble
of hydrogen atoms with the following frequencies:
\begin{equation}
\omega_{n,1} = (E_n-E_1)/\hbar .
\end{equation}
As discussed above, these photons are the consequences of the
quantization rule (34), breaking the Einstein's equation for
energy and passive gravitational mass. In the experiment we have
to use a macroscopic ensemble of hydrogen atoms to make the number
of the emitted photons to be large. More specifically, a tank of a
pressurized hydrogen is located in small spacecraft or satellite
and moved from a distant place, where gravitational potential is
small, $|\phi(R)| \ll |\phi(R_0)|$ , with constant velocity, $u
\ll \alpha c$, towards the Earth. Note that the latter inequality
allows us to disregard additional velocity dependent corrections
to the Hamiltonian (7),(8), which can be derived from the
Lagrangian of Ref. [3]. It is also important that each hydrogen
atom is at rest with respect to the spacecraft (satellite), which
means that gravitational force is compensated by some forces of
non-gravitational nature. Note that this changes a little the
allowed frequencies (39). Nevertheless, the latter effect is out
of our current consideration, since it is possible to show [16]
that the changes of the frequencies (39) are less than the
existing accuracies of their measurements. Other words, each
hydrogen atom does not feel gravitational acceleration, $\bf g$,
but rather feels time dependent gravitational potential,
$\phi(R-ut)$. Therefore, each hydrogen atom is affected by the
following time dependent Hamiltonian:
\begin{equation}
\hat H= \frac{\hat {\bf p}^2}{2m_e} -\frac{e^2}{r} +
\frac{\phi(R-ut)-\phi(R)}{c^2} \biggl[ m_e + 3 \frac{\hat {\bf
p}^2}{2m_e} - 2 \frac{e^2}{r} \biggl] \ .
\end{equation}
[For more rigorous derivation of Eq.(40), see Ref.[6].]

\subsection{Photon emission and mass quantization}

Here, we describe the suggested realistic experiment in more
detail. We consider a hydrogen atom to be in its ground state, at
$t=0$, and located at distance $R$ from a center of the Earth,
where the gravitational potential is small. The wave function of a
ground state, corresponding to Hamiltonian (7),(8), can be written
as:
\begin{eqnarray}
\tilde{\Psi}_1(r,t) = (1- \phi/c^2)^{3/2} \Psi_1[(1-\phi /c^2)r] \
\nonumber\\
\exp[-im_ec^2(1+\phi /c^2) t /\hbar]
\nonumber\\
\exp[-iE_1(1+\phi /c^2)t/\hbar] \ ,
\end{eqnarray}
where $\phi=\phi(R)$. At arbitrary moment of time, $t>0$, electron
wave function and time dependent perturbation for the Hamiltonian
(7),(8) in inertial coordinate system, related to the spacecraft
(satellite),  can be expressed as:
\begin{eqnarray}
\tilde{\Psi}(r,t) = (1-\phi/c^2)^{3/2} \sum^{\infty}_{n=1}
\tilde{a}_n(t) \Psi_n[(1-\phi/c^2)r]
\nonumber\\ \exp[-im_ec^2(1+\phi/c^2)
t /\hbar]
\nonumber\\
 \exp[-iE_n(1+\phi/c^2)t/\hbar] ,
\end{eqnarray}
\begin{equation}
\hat U ({\bf r},R,t) =\frac{\phi(R-ut)-\phi(R)}{c^2}  \biggl(3
\frac{\hat {\bf p}^2}{2m_e}-2\frac{e^2}{r} \biggl) .
\end{equation}

Application of the time-dependent quantum mechanical perturbation theory
[9] gives the following solutions for functions $\tilde a_n(t)$ in Eq.(42):
\begin{eqnarray}
\tilde{a}_n(t)= -\frac{V_{n,1}}{\hbar \omega_{n,1}c^2}
\biggl\{[\phi(R-ut)-\phi(R)]  \exp(i \omega_{n,1}t)
\nonumber\\
 + \frac{u}{i
\omega_{n,1}} \int^t_0 \frac{d \phi(R+ut)}{dR} d[\exp(i
\omega_{n,1}t)]\biggl\}, \ n \neq 1 \ ,
\end{eqnarray}
where $V_{n,1}$ and $\omega_{n,1}$ are given by Eqs.(32),(39). It
is important that under the suggested experiment the following
inequality is obviously fulfilled:
\begin{equation}
u \ll \omega_{n,1} R \sim \alpha c \ (R_0/r_B) \sim 10^{13} c,
\end{equation}
therefore, we can disregard the second term in the amplitude (44):
\begin{equation}
\tilde{a}_n(t)= -\frac{V_{n,1}}{\hbar \omega_{n,1}c^2}
[\phi(R-ut)-\phi(R)]  \exp(i \omega_{n,1}t) \ , \ n \neq 1 .
\end{equation}
Since $|\phi(R)| \ll |\phi(R-ut)|$, we can write probabilities,
corresponding to amplitudes of Eq.(46), in the following way:
\begin{equation}
\tilde{P}_n(t)= \biggl( \frac{V_{n,1}}{\hbar \omega_{n,1}}
\biggl)^2 \frac{\phi(R-ut)^2}{c^4} = \biggl( \frac{V_{n,1}}{E_n
-E_1} \biggl)^2 \biggl[ \frac{\phi(R')}{c^2} \biggl]^2 ,
\end{equation}
where $R'=R-ut$. It is important that the probabilities (47)
depend only on gravitational potential , $\phi'=\phi(R')$, in
final position of a spacecraft (satellite). Moreover, they
coincide with the probabilities, obtained in both thought
experiments [see Eqs.(33) and (38)]. This allows us to clarify
their physical meaning. Indeed, since the probabilities (47),(33),
and (38) are equal, we can conclude that all photons, emitted by a
macroscopic ensemble of hydrogen atoms during the suggested
realistic experiment, correspond to the breakdown of the
Einstein's equation for passive gravitation mass due to
quantization of the mass (34). As we discussed above, the excited
levels spontaneously decay with time and, therefore, it is
possible to observe the quantization law (34) indirectly by
measuring electromagnetic radiation from a macroscopic ensemble of
the atoms. In this case, Eq.(47) gives probabilities that a
hydrogen atom emits a photon with frequencies (39) during the time
interval $t$. [We note that dipole matrix elements for $nS
\rightarrow 1S$ quantum transitions are zero. Nevertheless, the
corresponding photons can be emitted due to quadrupole effects.]

 Let us estimate the probabilities (47):
 \begin{equation}
\tilde{P}_n = \biggl( \frac{V_{n,1}}{E_n - E_1} \biggl)^2
 \frac{\phi^2(R')}{c^4}  \simeq  0.49 \times 10^{-18}
 \biggl( \frac{V_{n,1}}{E_n-E_1} \biggl)^2 ,
\end{equation}
where, in Eq.(48), we use the following  numerical values of the
Earth's mass, $M \simeq 6 \times 10^{24} kg$, and its radius, $R_0
\simeq 6.36 \times 10^6 m$. It is important that, although the
probabilities (47),(48) are small, the number of photons, $N$, emitted
by macroscopic ensemble of the atoms, can be large since the
factor $V^2_{n,1}/(E_n-E_1)^2$ is of the order of unity. For
instance, for 1000 moles of hydrogen atoms, $N$ is estimated as
\begin{equation}
N_{n,1} = 2.95 \times 10^{8} \biggl( \frac{V_{n,1}}{E_n-E_1}
\biggl)^2 , \ N_{2,1} = 0.9 \times 10^8 ,
\end{equation}
which can be experimentally detected, where $N_{n,1}$ stands for a
number of photons, emitted with frequency $\omega_{n,1} = (E_n
-E_1)/\hbar$ [see Eq.(39)].

\section{Active gravitational mass in classical physics}

Here, we introduce active gravitational mass for a classical model
of a hydrogen atom. Suppose that we have a heavy positively
charged particle (i.e. proton) with bare mass $m_p$ and light
negatively charged particle (i.e., electron) with bare mass $m_e$,
where $m_p \gg m_e$. At large distances, $R \gg r_B$, from the
atom, gravitational potential in the first approximation is
\begin{equation}
\phi(R)=-G \frac{m_p + m_e}{R} \ ,
\end{equation}
where we do not take into account kinetic and Coulomb potential
energies contributions. Since $m_p \gg m_e$, we below disregard
kinetic energy of proton and consider it as a center of mass of
the atom. The next step is to define how kinetic and
Coulomb potential energies of electron contribute to the electron
active gravitational mass. To be more specific, we define active
gravitational mass of the atom from gravitational potential acting
on a small test body at rest at distances much high than the
"size" of the atom, $r_B$. For simplicity, we prescribe potential
and kinetic energies to electron and, therefore, consider
corrections to electron gravitational mass. It is possible to show
from general theory of a weak gravitational field [1,13] that
gravitational potential in our case can be written as
\begin{equation}
\phi(R,t)=-G \frac{m_p + m_e}{R}- G \int \frac{\Delta
T^{kin}_{\alpha \alpha}(t,{\bf r})+ \Delta T^{pot}_{\alpha
\alpha}(t,{\bf r})}{c^2R} d^3 {\bf r} ,
\end{equation}
where $\Delta T^{kin}_{\alpha \beta}(t,{\bf r})$ and $\Delta
T^{pot}_{\alpha \beta}(t,{\bf r})$ are changes of stress-energy
tensor component $T_{\alpha beta}(t, {\bf r})$ due to kinetic
and Coulomb potential energies, respectively. [Note that in Eq.(51)
and everywhere below we disregard
the so-called retardation effects]. Therefore, in the second
approximation in $1/c^2$, active electron gravitational mass can
be written as:
\begin{equation}
m^g_a = m_e + \frac{1}{c^2} \int [\Delta T^{kin}_{\alpha
\alpha}(t,{\bf r}) + \Delta T^{pot}_{\alpha \alpha}(t,{\bf r})] d^3{\bf r}.
\end{equation}

Let us write the standard expression for stress-energy tensor of a
moving point particle without electrical charge [1,13]:
\begin{equation}
T^{\alpha \beta}({\bf r},t) = \frac{m v^{\alpha}(t)
v^{\beta}(t)}{\sqrt{1-v^2/c^2}} \ \delta^3[{\bf r}-{\bf }r_p(t)],
\end{equation}
where $v^{\alpha}$ is a four-velocity, $\delta^3[...]$ is the
three-dimentional Dirac's delta-function, ${\bf r_p}(t)$ is a
trajectory of the particle in three-dimensional space. It is easy to
show by means of Eq.(53) that at low enough velocity, $v \ll c$,
\begin{equation}
\Delta T^{kin}_{\alpha \alpha} = 3 \frac{mv^2}{2}.
\end{equation}
The standard expression for stress-energy tensor of
electromagnetic field [1] can be written as:
\begin{equation}
T_{em}^{\mu \nu} = \frac{1}{4 \pi} [F^{\mu \alpha} F^{\nu}_{\
\alpha} - \frac{1}{4} \eta^{\mu \nu} F_{\alpha \beta} F^{\alpha
\beta}],
\end{equation}
where $F^{\alpha \beta}$ is the so-called electromagnetic field
tensor, $\eta_{\alpha \beta}$ is metric of the Minkowski
spacetime. This expression can be significantly simplified in our
case, where only electrical field is present. As a result, we
obtain the following formula for change of the stress-energy
tensor in the presence of the Coulomb potential energy:
\begin{equation}
\Delta T^{pot}_{\alpha \alpha} = -2\frac{e^2}{r}.
\end{equation}
Therefore, the total electron active gravitational mass can be
written in the same way as the electron passive mass,
\begin{equation}
m^g_a = m_e + \biggl(\frac{m_e{\bf
v}^2}{2}-\frac{e^2}{r}\biggl)/c^2 + \biggl(2\frac{m_e
{\bf v}^2}{2}-\frac{e^2}{r}\biggl)/c^2,
\end{equation}
where the last term is the virial one. Since the virial term
changes with time, we come to the conclusion that active
gravitational mass of a classical body changes with time.
Nevertheless we can introduce averaged over time active
gravitational mass, which occurs to be equivalent to energy [3,4],
since the averaged over time virial term is zero:
\begin{eqnarray}
<m^g_a> &= m_e + \biggl< 2 \frac{\hat
{\bf p}^2}{2m_e}-\frac{e^2}{r} \biggl>_t /c^2 + \biggl< 2
\frac{m_e {\bf v}^2}{2}-\frac{e^2}{r}\biggl>_t /c^2
\nonumber\\
&=m_e + E/c^2,
\end{eqnarray}

\section{Equivalence of the expectation values of active gravitational
mass and energy for stationary states}

Here, we use the so-called semi-classical approach to the general
relativity [20,21], where the Einstein's gravitational equation
can be written as:
\begin{equation}
R_{\mu \nu} - \frac{1}{2}R g_{\mu \nu} = \frac{8 \pi G}{c^2}
\bigl<\hat T_{\mu \nu} \bigl> ,
\end{equation}
where the last term represents the expectation value of quantum
stress-energy operator of the matter. In our case, expression for
active gravitational mass (57) can be represented as the following
Hamiltonian,
\begin{equation}
m^g_a = m_e +\biggl(\frac{{\bf
p}^2}{2m_e}-\frac{e^2}{r}\biggl)/c^2 + \biggl(2\frac{{\bf
p}^2}{2m_e}-\frac{e^2}{r}\biggl)/c^2,
\end{equation}
which can be easily quantized:
\begin{equation}
\hat m^g_a = m_e +\biggl(\frac{{\bf \hat
p}^2}{2m_e}-\frac{e^2}{r}\biggl)/c^2 + \biggl(2\frac{{\bf
\hat p}^2}{2m_e}-\frac{e^2}{r}\biggl)/c^2.
\end{equation}
Therefore, in the framework of semi-classical theory of gravity,
the expectation value of active gravitational mass, corresponding
to macroscopic ensemble of the atoms with each of them being in
its ground state, is equal to:
\begin{eqnarray}
<\hat m^g_a> &= m_e +\biggl< \frac{{\bf \hat
p}^2}{2m_e}-\frac{e^2}{r}\biggl>/c^2 +
\biggl<2\frac{{\bf \hat p}^2}{2m_e}-\frac{e^2}{r}\biggl>/c^2
\nonumber\\
&= m_e + E_1/c^2.
\end{eqnarray}
Thus, we conclude that the expectation values of active
gravitational mass and energy are equivalent for stationary
quantum states.

\section{Inequivalence between active gravitational
mass and energy for superpositions of stationary states}

In the previous section we have established the equivalence for
the expectation values of active gravitational mass and energy for
stationary quantum states. Below, we study if such equivalence
survives for superpositions of stationary quantum states. As in
section IV, we consider the simplest superposition of the ground
and first excited s-wave states in a hydrogen atom, where
electron wave function can be written as:
\begin{equation}
\Psi_{1,2}(r,t) = \frac{1}{\sqrt{2}} \bigl[ \Psi_1(r) \exp(-iE_1t)
+ \Psi_2(r) \exp(-iE_2t) \bigl].
\end{equation}
As we discussed this before, the expectation value of energy for
the wave function (63) does not depend on time. Nevertheless, the
expectation value of active gravitational mass operator (61)
oscillates with time and has the following form:
\begin{equation}
<\hat m^g_a> = m_e + \frac{E_1+E_2}{2c^2}+ \frac{V_{1,2}}{c^2}
\cos \biggl[ \frac{(E_1-E_2)t}{\hbar} \biggl] ,
\end{equation}
which coincides with Eq.(24), describing time dependent
oscillations of passive gravitational mass, where matrix
element of the virial operator, $V_{1,2}$ is given by Eq.(25). As
we discussed before such oscillations are of a pure quantum origin
and do not have classical analogs. They directly demonstrate
inequivalence of active gravitational mass and energy at a
macroscopic level. Nevertheless, in the same way as in section IV,
we can introduce modified equivalence principle for the
expectation values of active gravitational mass and energy by
means of averaging of Eq.(64) over time:
\begin{equation}
<<\hat m^g_a>>_t = m_e + (E_1 + E_2)/2c^2 = E/c^2.
\end{equation}
As we stressed in section IV, such averaging procedure is
principally different from that in Eqs.(1),(58) and does not have
classical analogs.

\section*{Acknowledgements}

We are thankful to N.N. Bagmet, V.A. Belinski, Steven Carlip,
Douglas Singleton, and V.E. Zakharov for useful discussions. This
work was supported by the NSF under Grant DMR-1104512.

$^*$Also at: L.D. Landau Institute for Theoretical Physics, 2
Kosygina Street, Moscow 117334, Russia.

%%%%%%%%%%%%%%%%%%%%%%%%%%%%%%%%%%%%%%%%%%%%%%%%%%%%%%%%%%%%


\begin{references}
\bibitem{Lebed-2}
We pay attention that to calculate the Hamiltonian in a
linear with respect to a small parameter $\phi(R) / c^2$
approximation, we do not need to keep the terms of the order of
$[\phi(R)/c^2]^2$ in the metric (2), in contrast to the classical
problem about perihelion precession of Mercury's orbit [13].
\bibitem{Lebed-1}
Note that passive gravitational mass of a composite body, as a
measurable quantity, has to be represented in quantum mechanics by
some Hermitian operator. The most logical and straightforward way
is to define the passive gravitational mass operator to be
proportional to a weight operator in a weak gravitational field
[i.e., the operator which couples with gravitational potential in
a linear approximation in Eq.(6)]. This extends definitions of
gravitational mass in classical case [3,4] and  corresponds to the
so-called E\"{o}tv\"{o}s method to measure gravitational mass
[13].
\bibitem{Lebed-4}
Due to symmetry of our problem, we need to keep in Eq.(28) only wave
functions, corresponding to $nS$ quantum states.
\bibitem{Lebed-5}
Note that our choice of adiabatically switched on gravitational
potential (29) with $\lambda \rightarrow 0$ allows to avoid extra
velocity dependent terms (see, for example, the Largangian in Ref. [3]).
\end{references}
\end{document}